\documentclass[aps,prl,preprint,groupedaddress,a4paper,byrevtex,showpacs,showkeys]{revtex4}

\usepackage{graphicx}
\usepackage{amsmath}
\usepackage{amssymb}

\begin{document}


\title{Evidence of ion diffusion at room temperature in microcrystals of the
Bi$_{2}$Sr$_{2}$CaCu$_{2}$O$_{8+\delta}$ superconductor}



\author{M.~Truccato}
\email[Electronic address: ]{truccato@to.infn.it}
\affiliation{\textquotedblleft NIS\textquotedblright  Centre of Excellence and INFM UdR Torino Universit\`a - Dip. Fisica Sperimentale - Via P. Giuria 1, I-10125, Torino, Italy}

\author{C.~Lamberti}
\affiliation{\textquotedblleft NIS\textquotedblright  Centre of Excellence and INFM UdR Torino Universit\`a - Dip. Chimica I.F.M. - Via P. Giuria 7, I-10125, Torino, Italy}

\author{C.~Prestipino}
\affiliation{\textquotedblleft NIS\textquotedblright  Centre of Excellence and INFM UdR Torino Universit\`a - Dip. Chimica I.F.M. - Via P. Giuria 7, I-10125, Torino, Italy}

\author{A.~Agostino}
\affiliation{\textquotedblleft NIS\textquotedblright  Centre of Excellence and INFM UdR Torino Universit\`a - Dip. Chimica Generale ed Organica Applicata - C.$^{so}$ Massimo D'Azeglio 48, I-10125, Torino, Italy}



\begin{abstract}
We have studied Bi-2212 microcrystals aged at ambient conditions for 40 days.
Combined x-ray absorption near edge structure (XANES) and x-ray fluorescence (XRF) measurements 
with $\mu$m space resolution show both an increase of Cu$^{+}$
with respect to Cu$^{2+}$ and an enrichment in Cu \textit{vs} Bi and Sr cation content
near the sample edges in the \textit{b} axis direction. A parallel study on an electrically contacted sample
has indirectly detected the O loss observing both a resistivity increase and an increase in sample thickness
near the edges. We conclude that the O out-diffusion along the \textit{b} axis is
accompanied by Cu cation migration in the same direction.
\end{abstract}



\maketitle



Single crystals of the Bi$_{2}$Sr$_{2}$CaCu$_{2}$O$_{8+\delta}$ (Bi-2212) high-$T_{c}$ superconductor
can be grown in samples whose length ($\gtrsim 500$ $\mu$m) is much greater than both their width and their thickness
(whiskers). This kind of samples has recently attracted remarkable interest from the point of view of
both basic and applied physics. In fact, they have proved to be suitable systems for the study
of the excess conductivity above $T_{c}\ $ \cite{Truccato_2004} and of the transport properties along
the $c$ axis,\cite{Latyshev_1999} because of their
high quality and small sizes. Moreover, they are also good candidates
for the fabrication of microscopic electronic devices based on the intrinsic Josephson junction
stack structure.\cite{Inomata_2003}
Therefore, the check of the whisker homogeneity in terms of elemental and electronic structures
has become an important issue.
As far as bulk samples (i.e. with sizes greater than about $1000 \times 200 \times 10 \ \mu\text{m}^3$)
are concerned, it is well-known that the material properties can be tuned by annealing in proper
atmospheres. This procedure usually involves temperatures above 400$^{\circ}$C in order
to ensure homogeneous oxygen diffusion throughout the sample on a few hours time scale,\cite{Li_1994}
so that the O-diffusion length ($\lambda$) is much greater than the sample size ($\ell$).
On the other hand, it has already been shown that Bi-2212 whiskers can undergo a significant increase
in the in-plane resistivity when aged at room temperature on a two-year time scale,\cite{Truccato_02}
most likely because of the oxygen loss induced by the favorable comparison between $\ell$ and $\lambda$.

Variations in the O non-stoichiometry of Bi-2212 whiskers must be accompanied by a change of the Cu oxidation state
in order to guarantee the electrostatic neutrality of the crystal. If $\lambda$
is smaller or comparable with $\ell$, then a gradient in the O content,
and therefore in the average Cu oxidation state, is expected. As Cu$^{+}$ and Cu$^{2+}$ have
very different chemical behaviors, a significant local modification around Cu cations
is expected when they undergo the oxidation state variation. Due to its chemical selectivity,
x-ray absorption near edge structure (XANES) is the most suitable spectroscopic technique able
to investigate both the oxidation and the coordination state of a selected transition metal cation
inside a matrix.\cite{Lamberti_2002,Lamberti_2003,Retoux_1990,Qi_1992} As third generation synchrotron
facilities provide microfocus beamlines, the determination of the local oxidation state of Cu inside
aged Bi-2212 single crystals is possible at the $\mu$m-scale. With this aim
we have performed $\mu$-XANES experiments at the ID22 beamline of the ESRF synchrotron.\cite{ID22}
To verify whether the variation of the local Cu oxidation state, induced by O diffusion,
is accompanied by a cation migration or not, also x-ray fluorescence (XRF) has been measured with
the same spatial resolution of 1 $\mu$m (vertical) $\times$ 4 $\mu$m (horizontal).

In order to better appreciate the expected variation of the Cu oxidation state induced by O migration, samples were
grown 40 days before the synchrotron measurements and aged at ambient conditions. The samples were prepared
by the method of the oxygenation of melt quenched plates, resulting in crystals with typical dimensions
of 500 $\times$ 20
$\times$ 1 $\mu$m along the $a$, $b$ and $c$ axis, respectively. High quality single crystals
were selected at the optical microscope and either mounted on a glass capillary, for synchrotron measurements,
or electrically contacted by Ag thermal evaporation and
diffusion.\cite{Truccato_02} The single-phase character of
the samples was checked by standard four-probe measurement of the electrical resistance $R$ \emph{vs}
the temperature $T$. The typical $R(T)$ behavior is reported in the inset (a) of Fig.~\ref{Fig_1},
which shows $T_c$ = 79.1 K. At the end of this experiment the sample was kept at room temperature
for 12 h, then the $R(T)$ measurement was repeated, resulting in the scattered triangles reported in
inset (b) of Fig.~\ref{Fig_1}. To appreciate the small but significant shift of the $R(T)$ curve
$\Delta R/R \simeq 5 \cdot 10^{-3}$
only a limited $T$ interval has been reported, but the same $\Delta R/R$ value holds over the whole
measured $T$ range. The observed
increase of the resistance reflects the small, but indirectly measurable loss of O undergone by
the sample during the aging at ambient conditions. By repeating the $R(T)$ measurement
on the sample aged 40 days, the observed $\Delta R/R$ is $\simeq$30\%,
reflecting a large electronic rearrangement to accommodate the anionic vacancies.

At the synchrotron the sample has been mounted on a goniometric head with the $c$ axis parallel to the beam and the
$a$ axis forming an angle of 4.95$^{\circ}$ with respect to the vertical.
The exact angle has been determined a posteriori from micro x-ray diffraction ($\mu$-XRD) data (see below).
Owing to the huge difference in the crystal
size along the $a$ and $b$ directions, the non perfectly vertical geometry allowed us
to finely scan the sample along the $b$ direction by a simple vertical movement. Fig.~\ref{Fig_1} reports
the normalized
XANES spectra, collected in fluorescence mode with a sampling step of 0.5 eV across the Cu K-edge.
The vertical scanning explores 90 $\mu$m with a sampling step of 2 $\mu$m.
Doing so, the first spectrum is collected almost in the middle of the sample in the $b$ direction and
the last one is collected almost in the edge.
By moving from the central position to the crystal edge, we observe an increase of the white line intensity
(first resonance after the edge) of 10\% and a red shift of the edge higher than 2 eV. This indicates an important
gradient of the oxidation state of Cu along the $b$ direction. At the cystal edges, where the atomic O$^{2-}$
anions are supposed to recombine to give O$_2$ molecules leaving the samples (and their 4 electrons),
a significant Cu$^{+}$ enrichement has been obserevd. As the energy shift between the K-edge of pure Cu$^{+}$
and Cu$^{2+}$ model compounds is typically 5.0-6.5 eV \cite{Lamberti_2002,Lamberti_2003}, assuming that in the
central position of the crystal we still have an homogeneous population of Cu$^{2+}$ cations, a fraction of
about 30\% of Cu$^{+}$ is estimated at the edges. This experimental evidence suggests that in the O loss process
the rate determining step is the O$^{2-}$ migration to the crystal surface and not the O$_2$ recombination
at the surface. In this regard, also Qui et al.\cite{Qi_1992} have observed a variation of both edge position
and white line intensity between XANES spectra measured on an oxidized and on a reduced bulk Bi-2212
single crystal ($1 \times 1 \times 0.1 \ \text{mm}^3$ in size). However, the absence of a $\mu$m-focused beam
forced the authors to collect the XANES spectra over the whole single crystal volume.
As a consequence, the effects of the oxidation/reduction processes were averaged and the resulting differences
in the XANES spectra were significantly smaller (0.5 eV of edge shift) with respect to what observed here.

Finally, to verify whether the combined O migration and Cu$^{2+}$ reduction has any effect on the local cation
distribution, we have collected a 90 $\mu$m (vertical) $\times$ 32 $\mu$m (horizontal) 2D XRF map by acquiring
2 s/point with a spatial sampling of 2 $\mu$m (vertical) $\times$ 4 $\mu$m (horizontal). We used a Li-doped Si
detector perpendicular to the incident beam monochromatized at 17.3 KeV. The XRF resolution
has been estimated to be better than 0.2 KeV from the full width at half maximum of the elastic peak, 
in agreement with
the instrumental data sheet value of 160 eV. Together with the
Compton signal (at 15.5 KeV) also the Bi($L_{\alpha}$,$L_{\beta}$,$L_{\gamma}$) peaks, the Bi $M$ multiplet, the
Sr($K_{\alpha}$,$K_{\beta}$) peaks, the Cu($K_{\alpha}$,$K_{\beta}$) peaks and the Ca($K_{\alpha}$,$K_{\beta}$)
peaks have been detected. To avoid any systematic error related to the conversion of XRF counts into
chemical stoichiometry, we decided to analyze the $r$(Cu,Bi) = Cu($K_{\alpha}$)/Bi($L_{\beta}$),
$r$(Cu,Sr) = Cu($K_{\alpha}$)/Sr($K_{\alpha}$) and $r$(Bi,Sr) = Bi($L_{\beta}$)/Sr($K_{\alpha}$) count ratios.
By averaging these ratios in central and in lateral positions, the quantities $\langle r_c(A,B) \rangle$ and
$\langle r_l(A,B) \rangle$ ($A,B$ = Cu, Bi and Sr) were obtained, respectively. Being $\langle r(A,B)\rangle$
the value of the ratio averaged over the whole map, the following three inhomogeneity factors $f(A,B) =
[ \langle r_l(A,B) \rangle - \langle r_c(A,B)\rangle ] / \langle r(A,B)\rangle $ ($A,B$ = Cu, Bi and Sr)
were obtained: $f(\text{Cu,Bi})$ = 0.10, $f(\text{Cu,Sr})$ = 0.07 and $f(\text{Bi,Sr})$ = -0.05.
Fig.~\ref{Fig_2} reports the $r$(Cu,Bi) map over a fraction of the sampled region.
From this figure the lateral Cu enrichment is apparent.

The whisker used for the electrical characterization was mapped by atomic force microscopy (AFM) at the
end of the 40 days aging process and the result is shown in Fig.~\ref{Fig_3}. The crystal, exhibiting an
almost flat $a$-$b$ surface in the as grown condition, clearly shows an
increased thickness $\Delta z$ at the borders along the $b$ direction.
This is the consequence of the O depletion, which is known to induce an increase of the $c$ axis lattice parameter
\cite{Emmen_1992,Liang_2002} whose average over the crystal size along the $c$ axis direction
($\simeq$200 unit cells) has been clearly detected by AFM. The $c$ axis increase, located at the border
of the crystal along the $b$ direction, well correlates with the higher fraction of Cu$^{+}$ singled out in the same
region by $\mu$-XANES (Fig.~\ref{Fig_1}). These independent evidences well agree with the anisotropy
in the in-plane O diffusion coefficients for Bi-2212 \cite{Li_1994} that testifies a slower migration process
along the $b$ direction and therefore enhances the possibility to evidence a compositional gradient.
On a quantitative ground, the measured $\Delta z/z$ (corresponding to $\Delta c/c$) is $\simeq$4\%,
indicating that the average $c$ value on the $b$ borders is around 32 \AA. $\mu$-XRD measurements,
collected in transmission mode with an image plate, resulted in a $c$ axis value of 32.2 \AA $\ $for the $b$ border
region.

Summarizing, we have presented the first combined $\mu$-XANES and $\mu$-XRF study on a single crystal Bi-2212 sample.
Space resolution, in the $\mu$m range, allowed us to single out a gradient along the $b$ direction in the O content,
evidenced by a change in the Cu oxidation state. Parallel AFM investigation has revealed a similar topological
gradient in the $c$ axis, which is again a consequence of the O gradient.
It has turned out that these anionic, electronic and structural rearrangements are accompanied by a
cation migration producing a Cu enrichment at the border of the crystal.

We are indebted to R. Tucoulou and M. Drakopoulos for the technical support at ID22 beamline of the ESRF.


%



\newpage

\bibliography{diffusion}

\begin{thebibliography}{12}
\expandafter\ifx\csname natexlab\endcsname\relax\def\natexlab#1{#1}\fi
\expandafter\ifx\csname bibnamefont\endcsname\relax
  \def\bibnamefont#1{#1}\fi
\expandafter\ifx\csname bibfnamefont\endcsname\relax
  \def\bibfnamefont#1{#1}\fi
\expandafter\ifx\csname citenamefont\endcsname\relax
  \def\citenamefont#1{#1}\fi
\expandafter\ifx\csname url\endcsname\relax
  \def\url#1{\texttt{#1}}\fi
\expandafter\ifx\csname urlprefix\endcsname\relax\def\urlprefix{URL }\fi
\providecommand{\bibinfo}[2]{#2}
\providecommand{\eprint}[2][]{\url{#2}}

\bibitem[{\citenamefont{Truccato et~al.}(2004)\citenamefont{Truccato, Rinaudo,
  Causio, Paolini, Olivero, and Agostino}}]{Truccato_2004}
\bibinfo{author}{\bibfnamefont{M.}~\bibnamefont{Truccato}},
  \bibinfo{author}{\bibfnamefont{G.}~\bibnamefont{Rinaudo}},
  \bibinfo{author}{\bibfnamefont{A.}~\bibnamefont{Causio}},
  \bibinfo{author}{\bibfnamefont{C.}~\bibnamefont{Paolini}},
  \bibinfo{author}{\bibfnamefont{P.}~\bibnamefont{Olivero}}, \bibnamefont{and}
  \bibinfo{author}{\bibfnamefont{A.}~\bibnamefont{Agostino}},
  \bibinfo{journal}{cond-mat/} \textbf{\bibinfo{volume}{0409717}}
  (\bibinfo{year}{2004}).

\bibitem[{\citenamefont{Latyshev et~al.}(1999)\citenamefont{Latyshev,
  Yamashita, Bulaevskii, Graf, Balatsky, and Maley}}]{Latyshev_1999}
\bibinfo{author}{\bibfnamefont{Y.~I.} \bibnamefont{Latyshev}},
  \bibinfo{author}{\bibfnamefont{T.}~\bibnamefont{Yamashita}},
  \bibinfo{author}{\bibfnamefont{L.~N.} \bibnamefont{Bulaevskii}},
  \bibinfo{author}{\bibfnamefont{M.~J.} \bibnamefont{Graf}},
  \bibinfo{author}{\bibfnamefont{A.~V.} \bibnamefont{Balatsky}},
  \bibnamefont{and} \bibinfo{author}{\bibfnamefont{M.~P.} \bibnamefont{Maley}},
  \bibinfo{journal}{Phys. Rev. Lett.} \textbf{\bibinfo{volume}{82}},
  \bibinfo{pages}{5345} (\bibinfo{year}{1999}).

\bibitem[{\citenamefont{Inomata et~al.}(2003)\citenamefont{Inomata, Kawae,
  Nakajima, Kim, and Yamashita}}]{Inomata_2003}
\bibinfo{author}{\bibfnamefont{K.}~\bibnamefont{Inomata}},
  \bibinfo{author}{\bibfnamefont{T.}~\bibnamefont{Kawae}},
  \bibinfo{author}{\bibfnamefont{K.}~\bibnamefont{Nakajima}},
  \bibinfo{author}{\bibfnamefont{S.~J.} \bibnamefont{Kim}}, \bibnamefont{and}
  \bibinfo{author}{\bibfnamefont{T.}~\bibnamefont{Yamashita}},
  \bibinfo{journal}{Appl. Phys. Lett.} \textbf{\bibinfo{volume}{82}},
  \bibinfo{pages}{769} (\bibinfo{year}{2003}), \bibinfo{note}{and references
  therein}.

\bibitem[{\citenamefont{Li et~al.}(1994)\citenamefont{Li, Kes, Fu, Menovsky,
  and Franse}}]{Li_1994}
\bibinfo{author}{\bibfnamefont{T.~W.} \bibnamefont{Li}},
  \bibinfo{author}{\bibfnamefont{P.~H.} \bibnamefont{Kes}},
  \bibinfo{author}{\bibfnamefont{W.~T.} \bibnamefont{Fu}},
  \bibinfo{author}{\bibfnamefont{A.~A.} \bibnamefont{Menovsky}},
  \bibnamefont{and} \bibinfo{author}{\bibfnamefont{J.~J.~M.}
  \bibnamefont{Franse}}, \bibinfo{journal}{Physica C}
  \textbf{\bibinfo{volume}{224}}, \bibinfo{pages}{110} (\bibinfo{year}{1994}).

\bibitem[{\citenamefont{Truccato et~al.}(2002)\citenamefont{Truccato, Rinaudo,
  Manfredotti, Agostino, Benzi, Volpe, Paolini, and P.Olivero}}]{Truccato_02}
\bibinfo{author}{\bibfnamefont{M.}~\bibnamefont{Truccato}},
  \bibinfo{author}{\bibfnamefont{G.}~\bibnamefont{Rinaudo}},
  \bibinfo{author}{\bibfnamefont{C.}~\bibnamefont{Manfredotti}},
  \bibinfo{author}{\bibfnamefont{A.}~\bibnamefont{Agostino}},
  \bibinfo{author}{\bibfnamefont{P.}~\bibnamefont{Benzi}},
  \bibinfo{author}{\bibfnamefont{P.}~\bibnamefont{Volpe}},
  \bibinfo{author}{\bibfnamefont{C.}~\bibnamefont{Paolini}}, \bibnamefont{and}
  \bibinfo{author}{\bibnamefont{P.Olivero}}, \bibinfo{journal}{Supercond. Sci.
  Technol.} \textbf{\bibinfo{volume}{15}}, \bibinfo{pages}{1304}
  (\bibinfo{year}{2002}).

\bibitem[{\citenamefont{Lamberti et~al.}(2002)\citenamefont{Lamberti,
  Prestipino, Bonino, Capello, Bordiga, Spoto, Zecchina, Moreno, Cremaschi,
  Garilli et~al.}}]{Lamberti_2002}
\bibinfo{author}{\bibfnamefont{C.}~\bibnamefont{Lamberti}},
  \bibinfo{author}{\bibfnamefont{C.}~\bibnamefont{Prestipino}},
  \bibinfo{author}{\bibfnamefont{F.}~\bibnamefont{Bonino}},
  \bibinfo{author}{\bibfnamefont{L.}~\bibnamefont{Capello}},
  \bibinfo{author}{\bibfnamefont{S.}~\bibnamefont{Bordiga}},
  \bibinfo{author}{\bibfnamefont{G.}~\bibnamefont{Spoto}},
  \bibinfo{author}{\bibfnamefont{A.}~\bibnamefont{Zecchina}},
  \bibinfo{author}{\bibfnamefont{S.~D.} \bibnamefont{Moreno}},
  \bibinfo{author}{\bibfnamefont{B.}~\bibnamefont{Cremaschi}},
  \bibinfo{author}{\bibfnamefont{M.}~\bibnamefont{Garilli}},
  \bibnamefont{et~al.}, \bibinfo{journal}{Angew. Chem. Int. Ed.}
  \textbf{\bibinfo{volume}{41}}, \bibinfo{pages}{2341} (\bibinfo{year}{2002}).

\bibitem[{\citenamefont{Lamberti et~al.}(2003)\citenamefont{Lamberti, Bordiga,
  Bonino, Prestipino, Berlier, Capello, D'Acapito, i~Xamena, and
  Zecchina}}]{Lamberti_2003}
\bibinfo{author}{\bibfnamefont{C.}~\bibnamefont{Lamberti}},
  \bibinfo{author}{\bibfnamefont{S.}~\bibnamefont{Bordiga}},
  \bibinfo{author}{\bibfnamefont{F.}~\bibnamefont{Bonino}},
  \bibinfo{author}{\bibfnamefont{C.}~\bibnamefont{Prestipino}},
  \bibinfo{author}{\bibfnamefont{G.}~\bibnamefont{Berlier}},
  \bibinfo{author}{\bibfnamefont{L.}~\bibnamefont{Capello}},
  \bibinfo{author}{\bibfnamefont{F.}~\bibnamefont{D'Acapito}},
  \bibinfo{author}{\bibfnamefont{F.~X.~L.} \bibnamefont{i~Xamena}},
  \bibnamefont{and} \bibinfo{author}{\bibfnamefont{A.}~\bibnamefont{Zecchina}},
  \bibinfo{journal}{Phys. Chem. Chem. Phys.} \textbf{\bibinfo{volume}{5}},
  \bibinfo{pages}{4502} (\bibinfo{year}{2003}).

\bibitem[{\citenamefont{Retoux et~al.}(1990)\citenamefont{Retoux, Studer,
  Michel, Raveau, Fontaine, and Dartyge}}]{Retoux_1990}
\bibinfo{author}{\bibfnamefont{R.}~\bibnamefont{Retoux}},
  \bibinfo{author}{\bibfnamefont{F.}~\bibnamefont{Studer}},
  \bibinfo{author}{\bibfnamefont{C.}~\bibnamefont{Michel}},
  \bibinfo{author}{\bibfnamefont{B.}~\bibnamefont{Raveau}},
  \bibinfo{author}{\bibfnamefont{A.}~\bibnamefont{Fontaine}}, \bibnamefont{and}
  \bibinfo{author}{\bibfnamefont{E.}~\bibnamefont{Dartyge}},
  \bibinfo{journal}{Phys. Rev. B} \textbf{\bibinfo{volume}{41}},
  \bibinfo{pages}{193} (\bibinfo{year}{1990}).

\bibitem[{\citenamefont{Qi et~al.}(1992)\citenamefont{Qi, Ren, Gao, Lee, Soo,
  and Wang}}]{Qi_1992}
\bibinfo{author}{\bibfnamefont{M.}~\bibnamefont{Qi}},
  \bibinfo{author}{\bibfnamefont{Z.~F.} \bibnamefont{Ren}},
  \bibinfo{author}{\bibfnamefont{Y.}~\bibnamefont{Gao}},
  \bibinfo{author}{\bibfnamefont{P.}~\bibnamefont{Lee}},
  \bibinfo{author}{\bibfnamefont{Y.~L.} \bibnamefont{Soo}}, \bibnamefont{and}
  \bibinfo{author}{\bibfnamefont{J.~H.} \bibnamefont{Wang}},
  \bibinfo{journal}{Physica C} \textbf{\bibinfo{volume}{192}},
  \bibinfo{pages}{55} (\bibinfo{year}{1992}).

\bibitem[{\citenamefont{http://www.esrf.fr/UsersAndScience/Experiments/Imaging%
/ID22/BeamlineManual}()}]{ID22}
\bibinfo{author}{\bibnamefont{http://www.esrf.fr/UsersAndScience/Experiments/I%
maging/ID22/BeamlineManual}}.

\bibitem[{\citenamefont{Emmen et~al.}(1992)\citenamefont{Emmen, Lenczowski,
  Dalderop, and Brabers}}]{Emmen_1992}
\bibinfo{author}{\bibfnamefont{J.~H. P.~M.} \bibnamefont{Emmen}},
  \bibinfo{author}{\bibfnamefont{S.~K.~J.} \bibnamefont{Lenczowski}},
  \bibinfo{author}{\bibfnamefont{J.~H.~J.} \bibnamefont{Dalderop}},
  \bibnamefont{and} \bibinfo{author}{\bibfnamefont{V.~A.~M.}
  \bibnamefont{Brabers}}, \bibinfo{journal}{J. Cryst. Growth}
  \textbf{\bibinfo{volume}{118}}, \bibinfo{pages}{477} (\bibinfo{year}{1992}).

\bibitem[{\citenamefont{Liang et~al.}(2002)\citenamefont{Liang, Lin, Maljuk,
  and Yan}}]{Liang_2002}
\bibinfo{author}{\bibfnamefont{B.}~\bibnamefont{Liang}},
  \bibinfo{author}{\bibfnamefont{C.~T.} \bibnamefont{Lin}},
  \bibinfo{author}{\bibfnamefont{A.}~\bibnamefont{Maljuk}}, \bibnamefont{and}
  \bibinfo{author}{\bibfnamefont{Y.}~\bibnamefont{Yan}},
  \bibinfo{journal}{Physica C} \textbf{\bibinfo{volume}{366}},
  \bibinfo{pages}{254} (\bibinfo{year}{2002}).

\end{thebibliography}

\newpage

\section{Figure captions}

\vspace{1.5cm}

\begin{figure}
\includegraphics[angle=0,width=8.6cm]{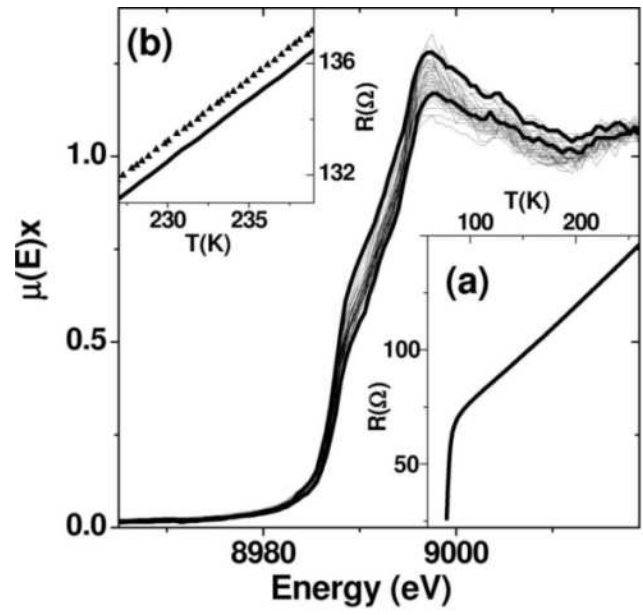}
\vspace{9.cm}
\caption{\label{Fig_1} Manuscript \# L05-0533 by Truccato \emph{et al}. To be reproduced in black and white.}
\end{figure}

Fig.~\ref{Fig_1}: Normalized XANES spectra collected on different positions of the Bi-2212 whisker.
A shift of the edge up to 2.5 eV, accompanied by a modification of the white line intensity clearly
testify an important modification of both the oxidation and the coordination state of Cu along the crystal.
Bold curves represent spectra collected in central and in near edge positions along the $b$ direction,
the others representing intermediate positions.
Inset (a) reports the $R$ \emph{vs} $T$ curve showing $T_c$ = 79.1 K. Inset (b)
compares, in a restricted $T$ range, the original measurement (solid line) with its repetition
after 12 h aging at 295 K (triangles).

\vspace{1.5cm}

\begin{figure}
\includegraphics[angle=0,width=8.8cm]{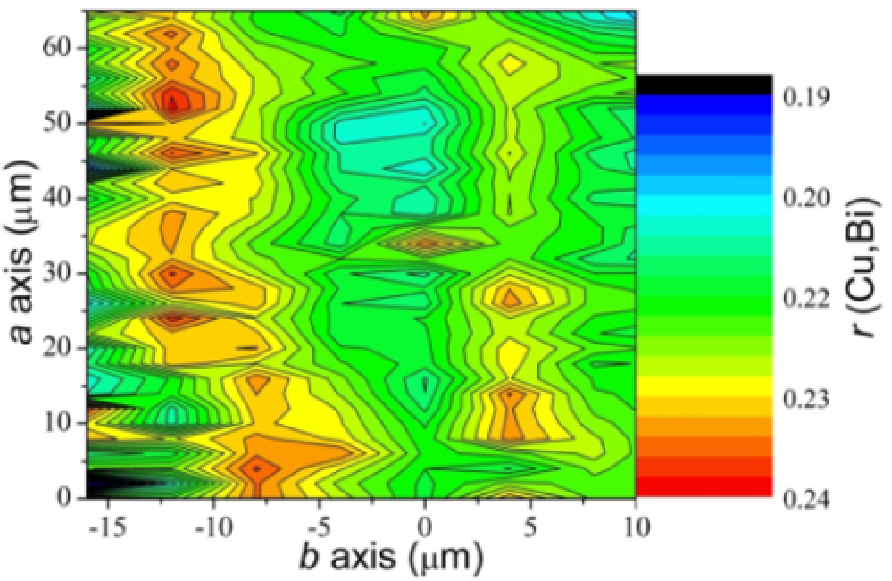}
\vspace{9.cm}
\caption{\label{Fig_2} Manuscript \# L05-0533 by Truccato \emph{et al}. To be reproduced in color both in the 
printed and in the online edition.}
\end{figure}

Fig.~\ref{Fig_2}: Color scale XRF map reporting the $r$(Cu,Bi) = Cu($K_{\alpha}$)/Bi($L_{\beta}$) ratio
on a portion of the same aged Bi-2212 sample used for the XANES study. The red regions correspond to the edge of the crystal.

\vspace{1.5cm}

\begin{figure}
\includegraphics[angle=0,width=8.9cm]{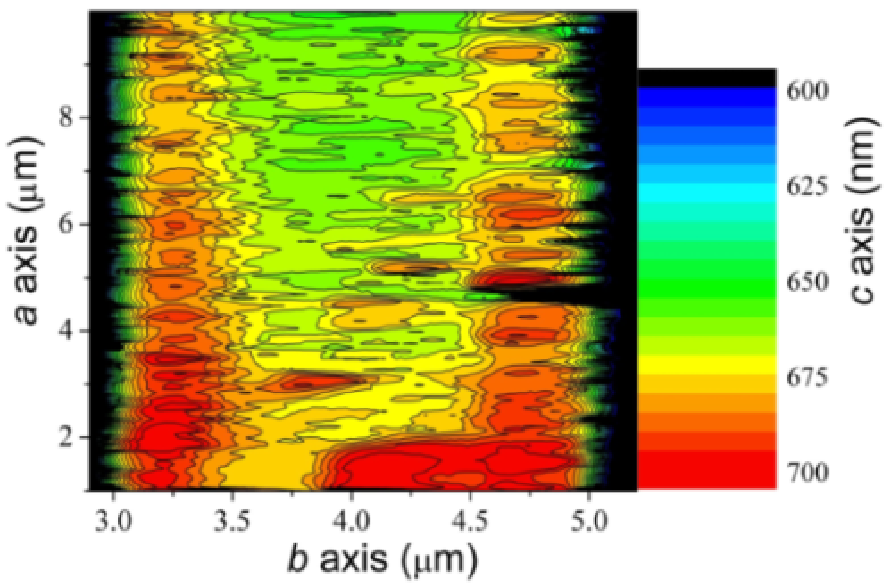}
\vspace{9.cm}
\caption{\label{Fig_3} Manuscript \# L05-0533 by Truccato \emph{et al}. To be reproduced in color both 
in the printed and in the online edition.}
\end{figure}

Fig.~\ref{Fig_3}: Color scale AFM map of the Bi-2212 sample used for the electrical measurements
at the end of 40 days aging process. The red regions correspond to the edge of the crystal.

\newpage

\newpage

\vspace{15.cm}
\newpage

\end{document}